\documentclass[conference]{IEEEtran}

\usepackage[utf8]{inputenc} 
\usepackage[T1]{fontenc}    
\usepackage{hyperref}       
\usepackage{url}            
\usepackage{booktabs}       
\usepackage{amsfonts}       
\usepackage{nicefrac}       
\usepackage{microtype}      
\usepackage{xcolor}         
\usepackage{caption}
\usepackage[rightcaption]{sidecap}
\usepackage{graphicx}
\usepackage{epstopdf}
\usepackage{multirow}
\usepackage[affil-it]{authblk}
\usepackage{amsmath,amssymb,amsfonts}

\usepackage[backend=biber]{biblatex}

\begin{document}

\title{ Using Test-Time Data Augmentation for Cross-Domain Atrial Fibrillation Detection from ECG Signals }

\author[1,*]{Majid Soleimani}
\author[2]{Maedeh H.Toosi}
\author[2]{Siamak Mohammadi}
\author[3]{Babak Hossein Khalaj}

\affil[1]{Department of Biomedical Engineering, McGill University, Montreal, Canada}
\affil[2]{School of Electrical and Computer Engineering, University of Tehran, Iran}
\affil[3]{Electrical Engineering Department, Sharif University of Technology}
\affil[*]{Corresponding author email: majid.soleimani@mail.mcgill.ca}

\maketitle

\begin{abstract}
  Atrial fibrillation (AF) detection from electrocardiogram (ECG) signals is crucial for early diagnosis and management of cardiovascular diseases. However, deploying robust AF detection models across different datasets with significant domain variations remains a challenge. In this paper, we use test-time data augmentation (TTA) to address the cross-domain problem and enhance AF detection performance. We use a publicly available dataset for training - Physionet Computing in Cardiology Challenge 2017 -, while collecting a distinct test set, creating a cross-domain scenario. We employ a neural network architecture that integrates transformer-based encoding of ECG signals and convolutional layers for spectrogram feature extraction. The model combines the latent representations obtained from both encoders to classify the input signals. By incorporating TTA during inference, we enhance the model's performance, achieving an F1 score of 76.6\% on our test set. Furthermore, our experiments demonstrate that the model becomes more resilient to perturbations in the input signal, enhancing its robustness. We show that TTA can be effective in addressing the cross-domain problem, where training and test data originate from disparate sources. This work contributes to advancing the field of AF detection in real-world scenarios.
\end{abstract}

\section{Introduction}
\label{sec:intro}

Integrating artificial intelligence (AI) into healthcare presents challenges due to variations in medical equipment, complexities of clinical environments, and the requirement for large training datasets~\cite{rajpurkar2022ai}. However, collecting such data in the medical field has some challenges. 
 Variations in medical equipment and the complexities of clinical environments can lead to single-source bias, where AI systems become overly tuned to specific training environments, reducing their effectiveness and generalizability across different settings\cite{hawkins2004problem}. Additionally, the quality and diversity of medical data are often insufficient due to issues like extensive data cleaning that does not accurately represent real-world complexities~\cite{rajpurkar2022ai}. In this context, domain adaptation emerges as a crucial technique in machine learning, addressing the issue of applying models trained on one domain (source) to another related but different domain (target) where labeled data may be scarce or absent. 

In our work, we tackled a cross-domain issue arising from the limited amount of ECG data available from wearable devices, which was insufficient for effective training. In order to address the lack of data, we used public datasets to train our models. However, we encountered challenges due to differences in recording settings, such as hardware and hospital equipment variations. These differences resulted in poor performance, underscoring the complexities of effectively implementing AI in diverse clinical environments\cite{lasko2024probabilistic}.

Recent studies have addressed the challenges of different domains through several methods. These include Few-Shot learning ~\cite{oh2022understanding,hu2022pushing}, learn latent feature across domains with Generative Adversarial Network (GAN) ~\cite{li2018domain,tzeng2017adversarial}, Self-Supervised Learning ~\cite{ericsson2021well,yue2021prototypical}. In our work we utilize Test time data augmentation (TTA) method. TTA was successfully employed in deep learning, in tasks like ECG signal classification, to improve model performance during inference. Unlike the traditional data augmentation applications during training, TTA focuses on augmenting input data at test time to enhance model robustness and generalization. In the context of ECG signal classification, TTA involves generating multiple augmented versions of the input ECG signal using transformations such as random cropping, rotation, scaling, or adding noise and then averaging the predictions of the model across these augmented versions. By incorporating diverse augmented samples during inference, TTA helps the model to make more reliable predictions by reducing the impact of noise and variability in the input signals. \cite{cubuk2019autoaugment}

It has been demonstrated that TTA is effective in improving the performance of deep learning models for ECG signal classification. For instance, in a study by \cite{hannun2019cardiologist}, TTA was applied to a deep neural network for detecting abnormal findings in ECG signals, resulting in improved classification accuracy. In a recent work by \cite{yildirim2018arrhythmia}, TTA was shown to enhance the robustness of a deep learning model for arrhythmia detection by considering multiple augmented versions of input ECG signals during inference. These findings highlight the potential of TTA as a valuable technique for enhancing the reliability and generalization of deep learning models in ECG signal analysis tasks.


This paper presents tree key contributions in the application-based wearable system design for AF detection as follows:

\begin{itemize}
    \item Our work presents a method specifically designed to effectively detect Atrial Fibrillation (AF), the world’s most common sustained arrhythmia, particularly in scenarios where initial labeled data is limited.
    \item We demonstrate that test-time data augmentation is an effective technique for addressing the cross-domain problem in ECG signal classification.
    \item We conduct a comparative analysis between test-time data augmentation and other machine learning methods, showing that test-time data augmentation yields comparable or superior results.
    
\end{itemize}
\section{Methods}

\subsection{Theory}

Various factors affect the process of data acquisition. For example, noise, bias, or disconnections in the data acquisition tool, missing data, etc. To model the uncertainty caused by the data acquisition process, we consider transformations with the following formula:
\begin{equation} \label{eq:TB}
X = T_B(X_0) + e,
\end{equation}

where $X_0$ is the real signal, $T$ is the transformation applied on $X_0$, $B$ is the transformation parameters, and $e$ is the noise of the data acquisition. $X$ is the recorded signal that is used for inference. If $T$ is a reversible transformation, and $T_B^{-1}$ is the reverse of $T_B$, then

\begin{equation} \label{eq:TBinv}
X_0 = T_B^{-1} (X - e).
\end{equation}

Noise usually has a Gaussian distribution,

\begin{equation} \label{eq:noise}
e = N(\mu,\sigma),
\end{equation}

where $\mu$ and $\sigma$ are the mean and standard deviation, respectively. If $Y$ and $Y_0$ are the target predictions for $X$ and $X_0$, respectively, then for classification we have $Y = Y_0$. Assuming $f(.)$ is the neural network and $\theta$ is the parameters learned during training, then
\begin{equation} \label{eq:nn}
Y = f(\theta, X).
\end{equation}

For classification, $Y$ is a discrete variable. Because $X$ is only one of the many possible observations of $X_0$, inference with $X$ can result in biased predictions. Therefore, we try to do inference with $X_0$ instead. Considering \ref{eq:TB}, \ref{eq:TBinv}, and \ref{eq:nn} we have
\begin{equation} \label{eq:long1}
Y = T_B(Y_0) = T_B(f(\theta, X_0)=T_B(f(\theta, T_B^{-1}(X-e))).
\end{equation}

Assuming that we know $p(B)$ and $p(e)$ as the probability distributions of $B$ and $e$, respectively, we can find a probability distribution for $Y$ 


\begin{equation} \label{eq:long2}
p(Y|X) = p(T_B(f(\theta, T_B^{-1}(X-e)))), \text{ where } B \sim p(B), e \sim p(e).
\end{equation}

Considering the complexity of the above equation, we should estimate the final inference with Monte Carlo simulation. Monte Carlo simulation involves generating multiple random samples from probability distributions that represent uncertainties in the data or model parameters. These samples are then used to perform repeated simulations or calculations, allowing for the exploration of possible outcomes and their associated probabilities. \cite{metropolis1949monte}

With Monte Carlo simulation runs, the inference of the $n$'th simulation is

\begin{equation} \label{eq:long4}
y_n = T_{B_n}(f(\theta, T_{B_n}^{-1}(X-e_n)))).
\end{equation}

For calculating $y_n$, we should first sample $B_n$ and $e_n$ from $p(B)$ and $p(e)$, respectively, resulting in $T_{B_n}$, and then we use \ref{eq:long1}. Using $N$ Monte Carlo simulation runs a set of classification predictions is achieved,

\begin{equation} \label{eq:Y}
\mathcal{Y} = \{ y_1, y_2, y_3, ...y_N \}.
\end{equation}
Finally, we can estimate the final inference


\begin{equation} \label{eq:yhat2}
\hat{Y} = \arg \max_{y} p(y|X) \approx \text{Mode}(\mathcal{Y}),
\end{equation}

where $\text{Mode}(\mathcal{Y})$ is the element in $\mathcal{Y}$ with the maximum repetition. \cite{wang2019aleatoric}

\subsubsection{Neural network structure}

In this study, we used the ECG-DualNet ++ network introduced in \cite{rohr2022exploring}. This network offers several advantages in the domain of ECG signal classification:

\begin{itemize}
    \item Out-of-Distribution Robustness: This neural network has been evaluated in an out-of-distribution setting, wherein the test set originates from a distribution distinct from the training set, demonstrating its robustness and generalizability.
    \item Comparable Performance: The network's performance is on par with the top 15 deep learning models, highlighting its competitive efficacy for ECG signal classification.
    \item Automated Feature Extraction: Unlike the networks with the best results on PhysioNet that typically rely on hand-crafted features derived from ECG medical information, this network exclusively employs deep learning features, eliminating the need for human review and manual feature extraction.
\end{itemize}

\begin{figure*}[ht]
   \centering
       \includegraphics[width=0.9\textwidth]{"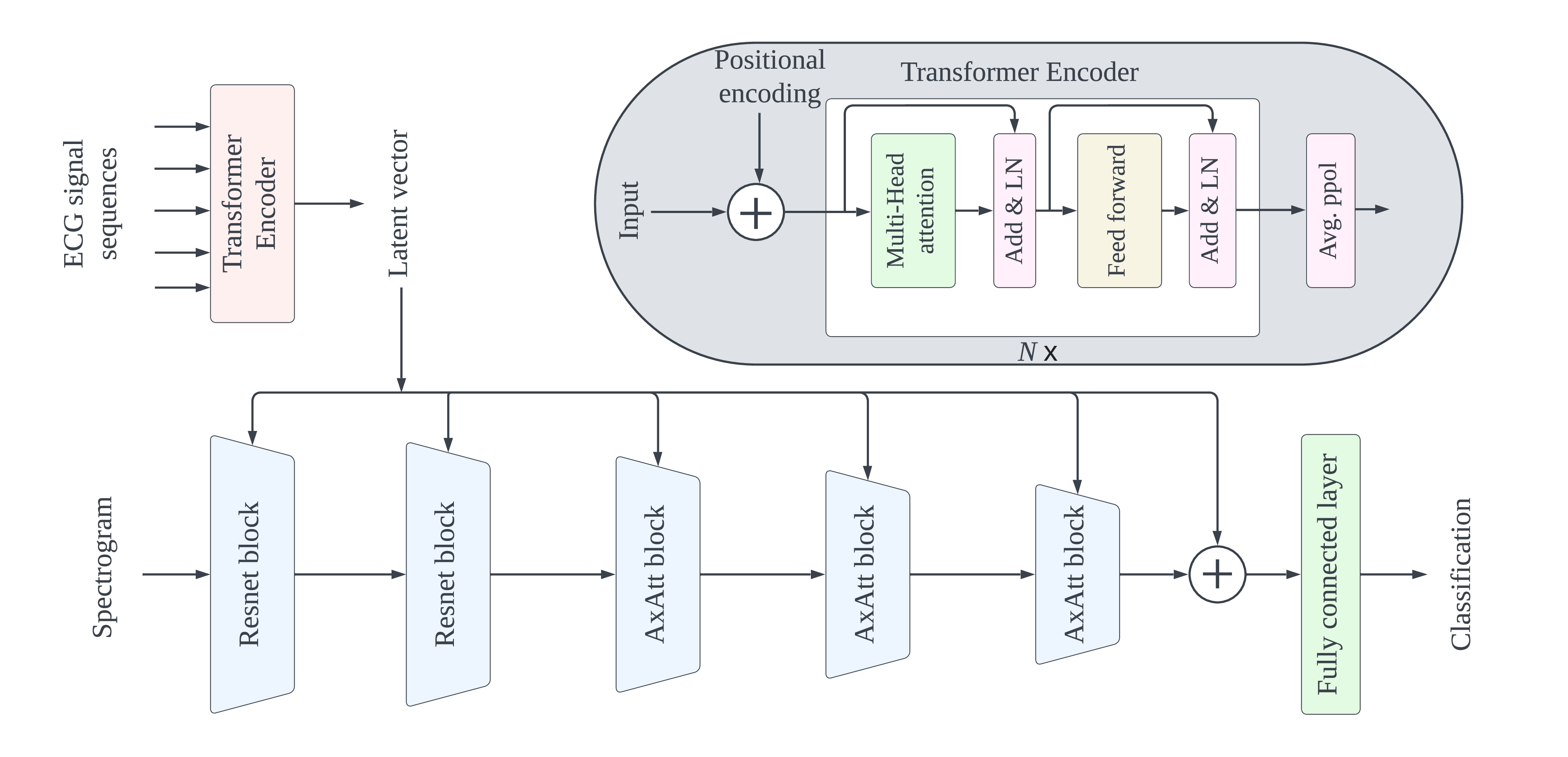"} 
       
 \caption{The neural network structure \cite{rohr2022exploring}. The ECG signal is fed into the transformer encoder that gives the latent vector. The spectrogram of the same signal is fed into two residual blocks followed by three Axial-Attention blocks. All of the spectrogram encoder blocks also include the latent vector with a conditional batch normalization. Finally, the two encoded vectors are added to each other and the classification is done by a fully connected layer with softmax activation.}
 \label{fig:NN}
\end{figure*}

The ECG-DualNet(++) architecture is completely explained in \cite{rohr2022exploring}.
The ECG-DualNet(++) architecture is depicted in Figure \ref{fig:NN}. As seen in the figure, it consists of two main parts: a signal encoder and a spectrogram encoder. 

The input to the signal encoder is short sequences derived from the time-domain ECG signal. This input is utilized to predict a latent vector. The signal encoder encodes the signal by a transformer encoder. The transformer encoder architecture is a simple transformer proposed by \cite{vaswani2017attention}, which is a stack of $N$ identical blocks. Each block has two sub-blocks: a multi-head self-attention block and a fully connected feed-forward block. The output of each block is added to the input to that block and then is normalized. 

The spectrogram encoder processes the spectrogram of the ECG signal. The input to this branch of the network is the ECG signal spectrogram. The spectrogram encoder reduces the dimensionality of frequency-domain features within each block. This encoder consists of two simple residual blocks \cite{he2016deep} and three axial attention blocks which ensure both large receptive fields and efficient computation costs \cite{wang2020axial}.

In order to combine the two encoders, a special batch normalization algorithm (conditional batch normalization) is integrated into each spectrogram encoder block as per \cite{de2017modulating}, merges the latent vector from the signal encoder with the frequency-domain features. 

At the end, the classification label is predicted by a fully connected layer with softmax activation

\section{Experiments}

\subsection{datasets}
Our proposed model is developed using ECG signals derived from two distinct datasets. The training dataset, which includes all ECG data from the public Physionet Computing in Cardiology Challenge 2017~\cite{clifford2017af}, consists of 8,482 single-lead ECG recordings classified into four classes: normal, AF, other rhythms, and noisy. For testing, we use a private dataset collected from a wearable device comprising 629 two-lead ECG recordings containing normal and AF rhythms.
Additional details about these datasets can be found in Table~\ref{tab:dataset-description}. The following section outlines the methodology used to prepare the model.

\subsection{Data Augmentation}

\begin{figure*}[t]
  \centering
  \includegraphics[width=\linewidth]{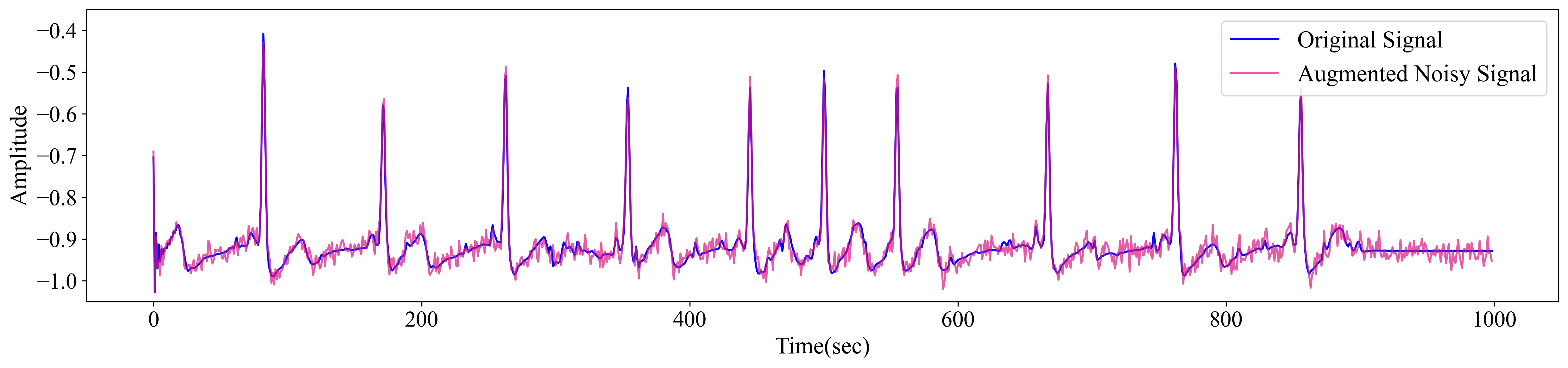}
  \caption{Original signal with noise augmentation method}
  \label{fig:signal_noise}  
\end{figure*}

Dealing with an imbalanced dataset is proving to be a challenging task during our training process. The normal class has a much higher number of records, with 5,154 ECG records, compared to only 771 records for the AF class. This non-uniform distribution can cause overfitting, making it difficult to train and evaluate the model effectively.
To address this issue, we employ two separate data augmentation methods. The first method combines random resampling with SMOTE (Synthetic Minority Over-sampling Technique)\cite{hospedales2021meta}. Initially, random resampling corrects the imbalance between the AF and normal classes by adjusting the number of samples. Then, SMOTE is used to generate synthetic data points by interpolating between existing samples in the minority class's feature space. This technique enhances the diversity of the dataset and helps smooth class boundaries, reducing the risk of overfitting by introducing more varied examples.
The second method uses a combination of Gaussian noise and random resampling. In this approach, Gaussian noise is added to the data points after adjusting their quantity through random resampling, as shown in Figure ~\ref{fig:signal_noise}. This simulates real-world measurement errors and helps the model learn to generalize better by becoming less sensitive to minor variations in input data, thus enhancing its robustness in diverse operational conditions.

\subsubsection{Data augmentation methods for training and TTA}

To improve the robustness of the model, we use various data augmentation methods during training. A random combination of these methods is created for each sample during training, while the parameters of each method are also randomly chosen. The same list of data augmentation methods was used for TTA.

The methods used for on-the-fly data augmentation are:
\begin{itemize}
    \item Drop: randomly missing signal values
    \item Mask: randomly missing an interval signal
    \item Shift: randomly shift the signal to the right with zero padding to the left.
    \item Sine: add a random sine wave to the entire signal
    \item Band-pass filter: aplly a band-pass filter to the signal
    \item Cut-mix: randomly cut and mix parts of signals from the same class.
    \item Random flip: randomly flip the signal along the voltage axis
    \item Noise: add a normal noise to the signal with random signal-to-noise ration ($SNR$) 
\end{itemize}

\subsection {Data Pre-processing}

After acquiring the ECG signals from the datasets, preprocessing was performed to enhance signal quality and remove noise. The first step involved the application of a median filter to mitigate baseline wander, a common artifact in ECG recordings. Subsequently, a second-order band-pass Butterworth filter was applied to filter out unwanted frequency components and retain the signal of interest within the desired frequency range. Following this, downsampling at 100 Hz was applied using interpolation techniques to ensure uniformity in sampling rates across different datasets. The signals were then normalized using z-score normalization to facilitate consistent analysis and interpretation.

\begin{table}[t]
    \centering
    \caption{Summary of characteristics for the train and test datasets used in the study}
    \label{tab:dataset-description} 
    \begin{tabular}{|c|c|c|c|c|}
    \hline
    Dataset & Label & Label  & Frequency & Duration \\
     & (Normal) & (AF)  & [MHz] & [sec] \\
    \hline
    Train & 5154 & 771 & 300 & 9-60 \\
    \hline
    Test & 332 & 297 & 200 & 40 \\
    \hline
    \end{tabular}
\end{table}

\label{sec:exp}
\begin{figure*}[t] 
  \centering  
  \begin{minipage}[t]{0.47\textwidth} 
    \centering
    \includegraphics[width=0.8\linewidth]{"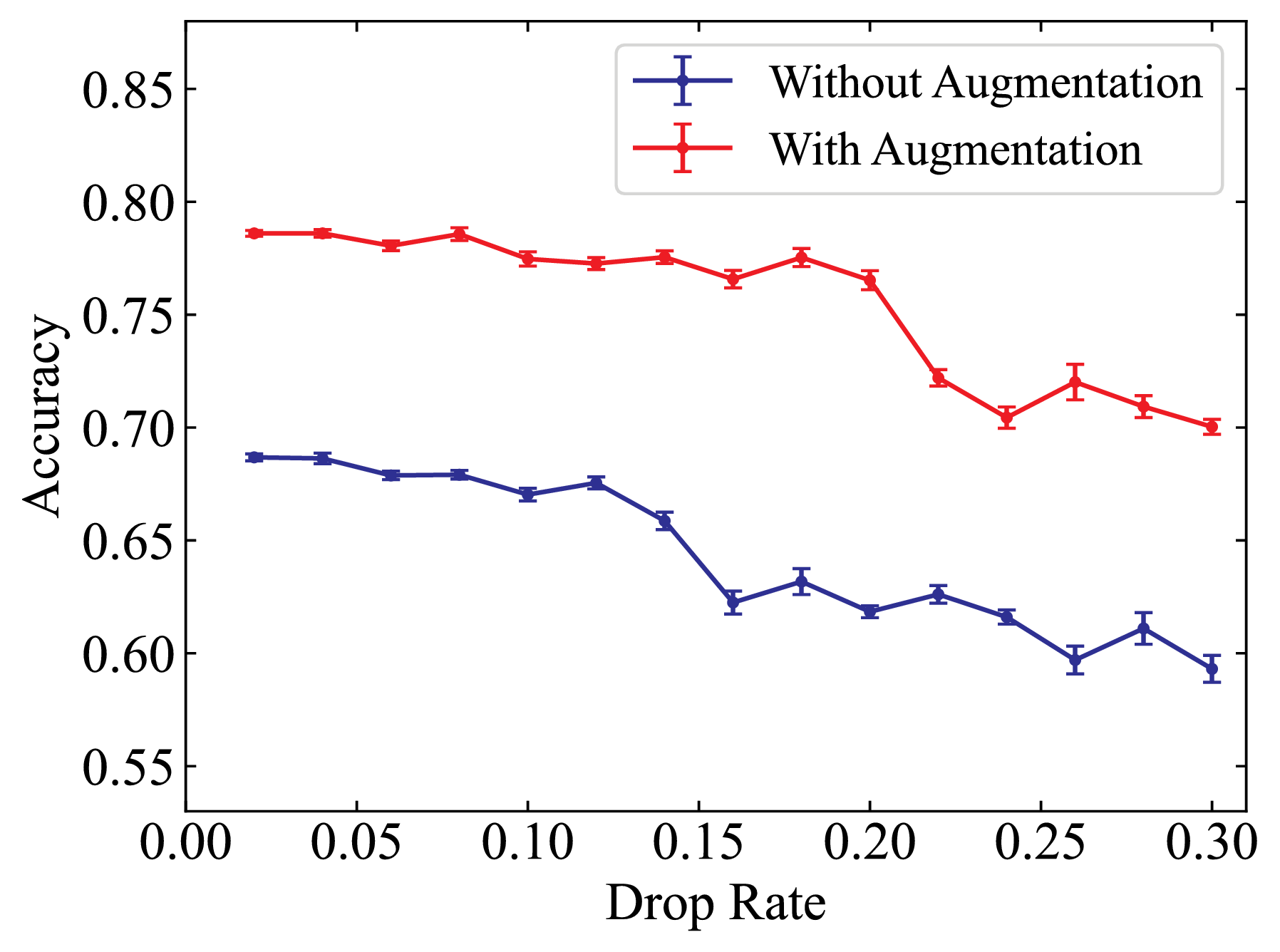"} 
    \caption{The network performance when increasing the drop rate by setting ECG values to zero during testing shows that accuracy decreases more and at lower drop rates when the network is trained without data augmentation (blue line) compared to with data augmentation (red line). Error bars represent standard deviations from 10 experiment repetitions.} 
    \label{fig:Drop} 
  \end{minipage}
  \hfill 
  \begin{minipage}[t]{0.47\textwidth} 
    \centering
    \includegraphics[width=0.8\linewidth]{"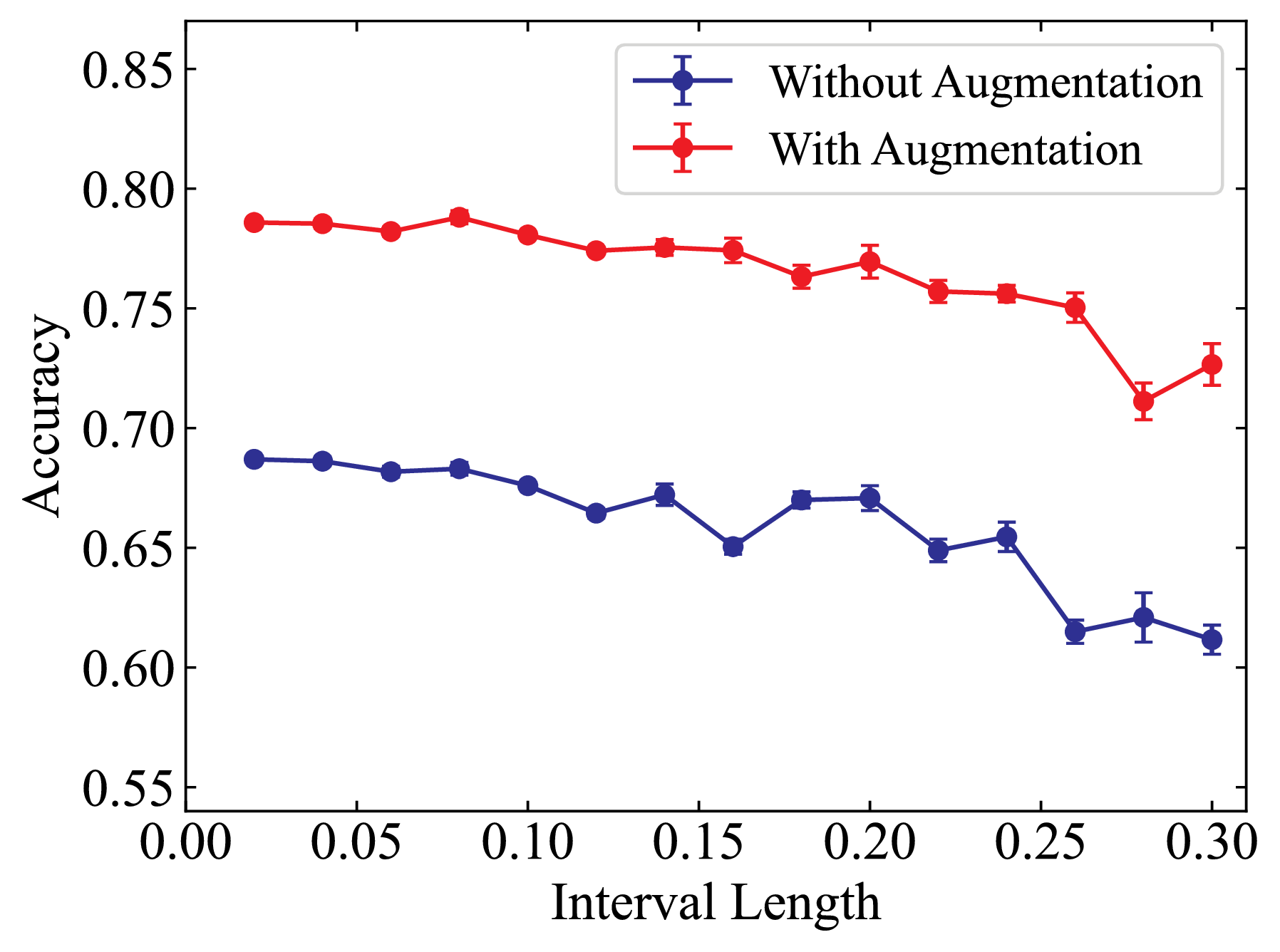"} 
    \caption{The network performance with increasing missed interval lengths by setting ECG values to zero during testing shows that accuracy decreases more and at shorter intervals when the network is trained without data augmentation (blue line) compared to with data augmentation (red line). Error bars indicate standard deviations from 10 experiment repetitions.} 
    \label{fig:Mask}
  \end{minipage}
\end{figure*}

\begin{figure}[t] 
  \centering  
  \begin{minipage}[t]{0.47\textwidth} 
    \centering
    \includegraphics[width=0.9\linewidth]{"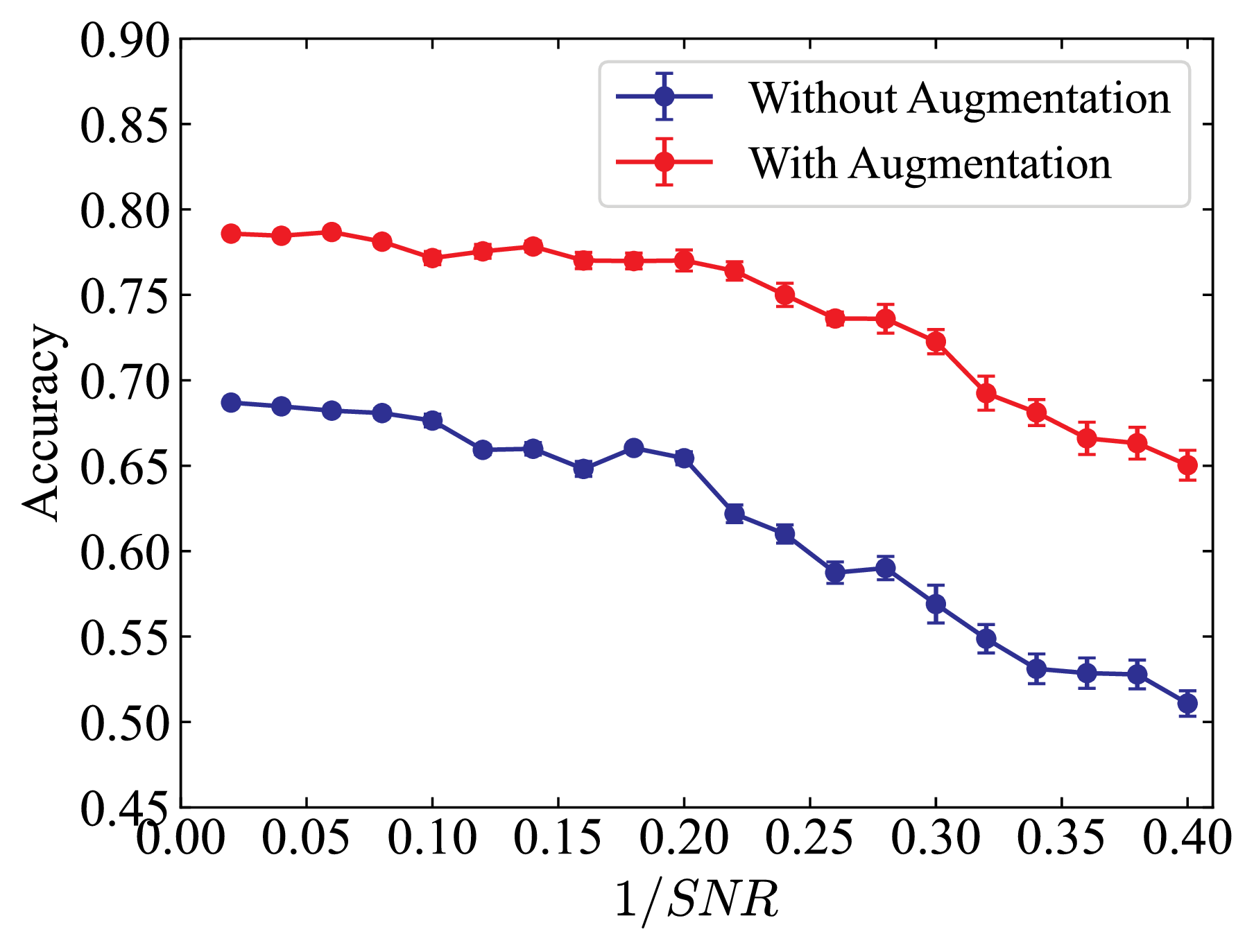"} 
    \caption{The summary of the network performance with decreasing signal-to-noise ratio (SNR) by adding random noise to the ECG signal during testing shows that accuracy decreases more and at higher SNRs when the network is trained without data augmentation (blue line) compared to with data augmentation (red line). Error bars represent standard deviations from 10 experiment repetitions.} 
    \label{fig:noise} 
  \end{minipage}
  \hfill 
  \begin{minipage}[t]{0.47\textwidth} 
    \centering
    \includegraphics[width=0.9\linewidth]{"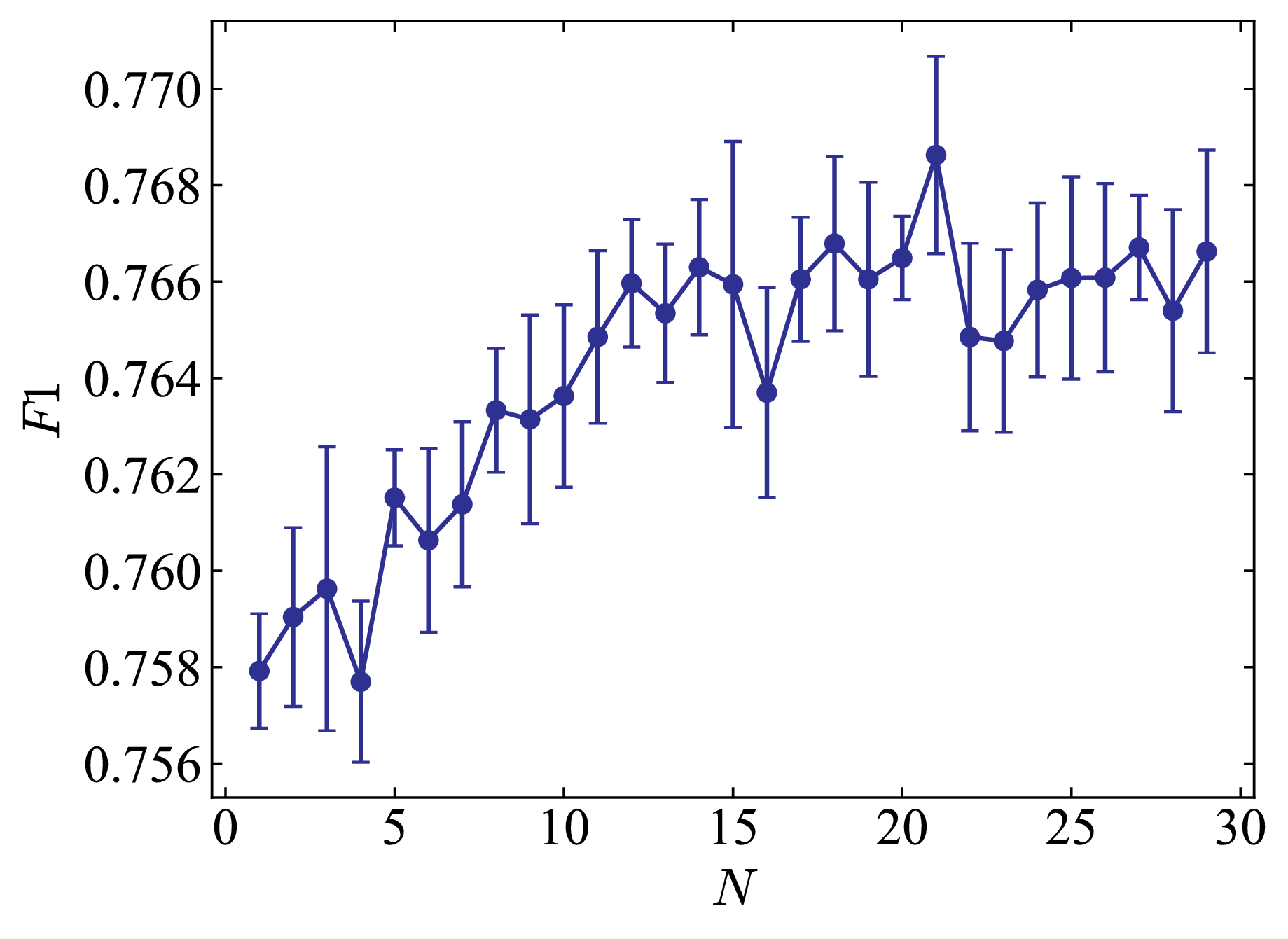"} 
    \caption{Improvements in the F1 score with TTA as $N$ is increased, that is the number of Monte Carlo simulation runs. The error-bars are the standard deviations for repeating the experiments 10 times.} 
    \label{fig:TTA}
  \end{minipage}
  \vspace{-1.5em}
\end{figure}

\subsection{Model robustness with TTA}
We first explore the impact of data augmentation intensity on the accuracy of neural network inference for ECG signal classification, utilizing three distinct methods: drop, mask, and noise. The drop method randomly sets ECG values to zero (figure \ref{fig:Drop}), the mask method randomly sets intervals of the ECG signal to zero (figure \ref{fig:Mask}), and the noise method adds random Gaussian noise to the ECG signal (figure \ref{fig:noise}). Our results show that without data augmentation during training, accuracy decreases more and at lower augmentation intensities compared to when data augmentation is employed.

We repeated each test 10 times to ensure reliability, with the standard deviation (STD) of results calculated for each perturbation intensity level. Figures \ref{fig:Drop}, \ref{fig:Mask}, and \ref{fig:noise} show that standard deviation, depicted as error bars, increases with the rise in perturbation intensity. These diagrams reveal the effectiveness of each data augmentation method in mitigating accuracy degradation. While all methods improve accuracy compared to no data augmentation during training, the noise method exhibits a more pronounced decrease in accuracy. This suggests that adding random Gaussian noise introduces more disruptive perturbations, impacting model performance during inference.

Understanding the relative effectiveness of each data augmentation method offers insights into the specific weaknesses and vulnerabilities of the neural network model against different types of augmentation-induced perturbations. These results imply that strategic adjustments in training data augmentation can enhance model generalization. By leveraging insights from this comparative analysis, improved data augmentation pipelines can enhance model performance with real-world signals. This iterative process can improve the robustness and reliability of neural network models in ECG signal classification tasks, particularly against domain shifts and out-of-distribution samples.

Figure \ref{fig:TTA} demonstrates the efficacy of TTA in enhancing model performance, showing F1 score improvements with increasing Monte Carlo simulation runs. F1 scores improved by more than 1 percent with a standard deviation of 0.002\%. Despite inherent randomness, results reveal a trend towards stability in F1 scores beyond 15 augmented samples, highlighting the robustness of the TTA approach for ECG classification and ensuring consistent and reliable predictions in real-world scenarios with prevalent data variability.

\subsection{Comparison with the State-of-the-art Methods}
In this section, we will conduct a comparison of our work with existing state-of-the-art methods to demonstrate the superior performance of our method in contrast to them.

\begin{figure}[t]
    \centering
        \includegraphics[width=0.8\linewidth]{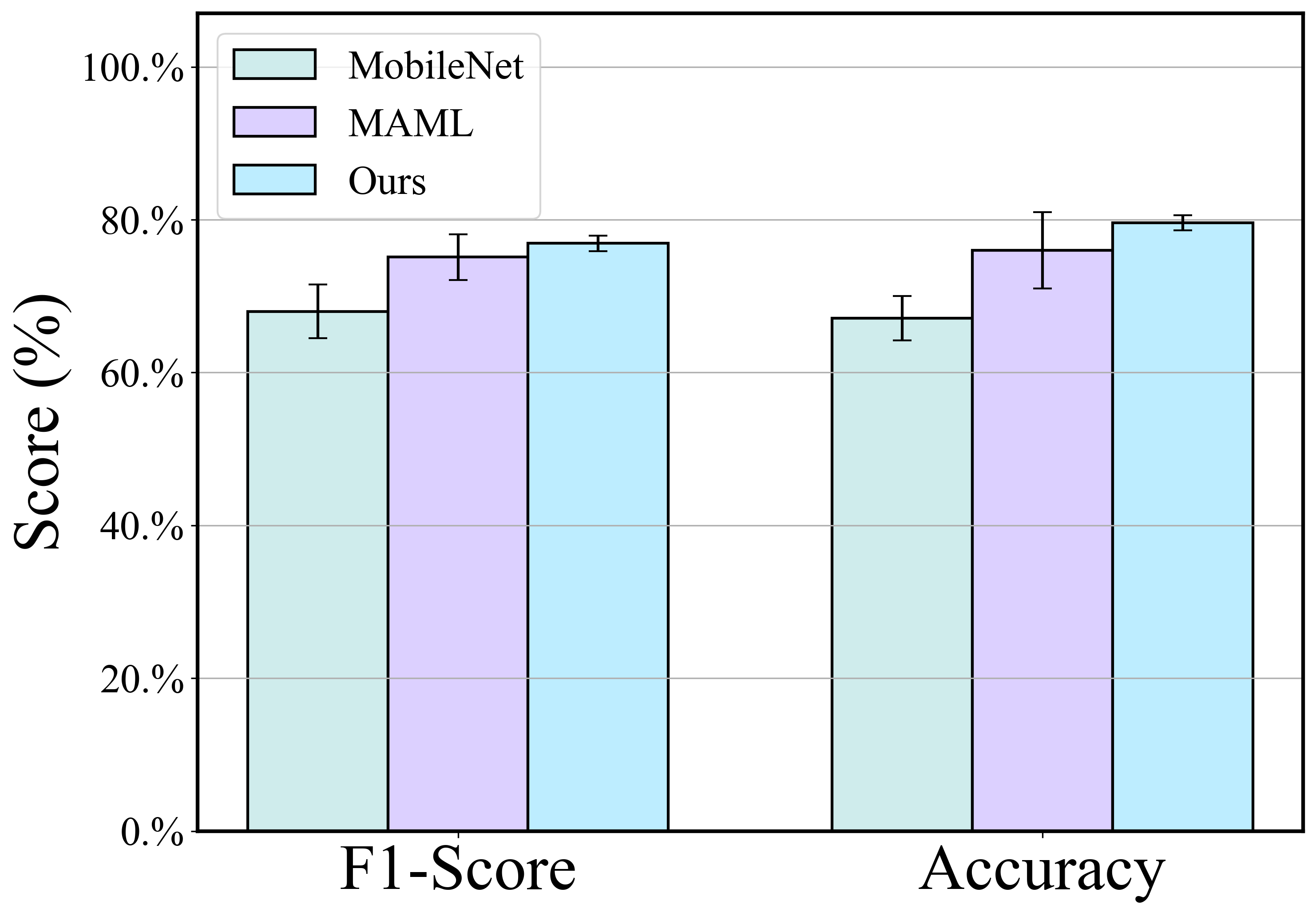}
        \caption{Comparison of the results between our method and the other State-of-the-art methods}
        \label{fig:comparison}
\end{figure}

\textbf{Deep Learning (DL)} has revolutionized healthcare by providing a transformative approach in detecting arrhythmias using electrocardiogram (ECG) signals. With the introduction of DL in this domain, more accurate, efficient, and real-time predictions and classifications of various heart conditions have become possible. DL techniques have proven to be highly effective in enhancing the automated analysis of ECG signals for arrhythmia detection. For instance, Convolutional Neural Networks (CNNs) have been extensively applied in state-of-the-art methods due to their superior capability in feature extraction. ~\cite{huang2019ecg,chen2024novel,atal2020arrhythmia}.Due to its efficiency and lightweight structure, we employ the MobileNet architecture proposed in ~\cite{kim2022lightweight} to detect AF arrhythmias. 
Fig ~\ref{fig:comparison} illustrates the effectiveness of our model over the MobileNet~\cite{kim2022lightweight} in the classification of AF, which shows significant enhancements in classification performance. Notably, our model outperforms MobileNet with an F1 score of 76.8\% or a rise of 9\%, and an accuracy rate of 79.6, or a rise of 13\%;

starting with an input layer that processes incoming ECG signals, followed by ten inverted residual blocks designed for efficient data processing. Each block uses a depthwise convolutional layer that expands its depth by six times and then reduces it back to the original channel size through a projection layer. The structure employs strides of two, and channel outputs progressively increase from 8 to 24 across the blocks. 


\textbf{Meta Learning}, also known as learning-to-learn, addresses the challenge of cross-domain classification by training models to be robust against domain shifts. This approach focuses on optimizing feature-wise transformation layers to tailor the distribution of features across various domains, enhancing adaptability to new, unseen domains(\cite{oh2022understanding,li2018learning,tseng2020cross}). Techniques such as Model-Agnostic Meta-Learning (MAML)\cite{finn2017model}, Prototypical Networks\cite{snell2017prototypical}, Siamese Networks\cite{koch2015siamese}, and matching\cite{vinyals2016matching} are particularly effective in this context. Notably, MAML, an optimization-based meta-learning algorithm introduced by ~\cite{finn2017model}, has shown promising performance in medical applications, demonstrating robustness across domain shifts\cite{zhu2022enhancing,liu2021metaphys}. 
This approach enhances the model's adaptability and generalization capabilities across various domains. 

Before discussing the implementation and results of MAML, it is essential to establish some foundational concepts related to meta-learning 
Unlike traditional supervised learning, which typically relies on a static dataset from a single task, meta-learning employs a dynamic, multi-task framework. A \textbf{task} refers to a specific learning challenge or problem, each with distinct characteristics and objectives, designed to train and evaluate the model’s adaptability and learning capabilities. The learning process consists of two main phases: \textbf{meta-training} and \textbf{meta-testing}. During meta-training, the model learns from multiple tasks, adapting its parameters to not only excel on a single task but also to generalize across various tasks. Meta-testing then assesses the model's ability to apply its learned knowledge to new, unseen tasks, thereby testing its generalization capabilities. Each task within these phases utilizes two distinct sets of data:

\textbf{Support Set:} Used for training the model on a particular task, this set includes a limited number of examples from which the model learns.

\textbf{Query Set:} Following training, this set tests the model, comprising different examples to evaluate how well the model can transfer its learned knowledge to new data.

In contexts such as classification, this approach is often termed \textbf{'K-shot, N-way'} classification, where the support set contains \textbf{'k'} instances for each of the \textbf{'N'} classes, blending principles from both meta-learning and few-shot learning.

MAML has two levels of optimization; it operates through a meta-training phase involving two distinct loops: an inner loop and an outer loop. During the inner loop, MAML updates its model parameters based on gradient calculations from training examples specific to each task. It then evaluates the model's performance by computing the loss using test examples from that same task. In the outer loop, MAML aggregates the losses from each task after these initial updates and applies a meta-gradient update to the original model parameters. This double-layered approach allows for refined adaptation across multiple tasks. During the meta-testing phase, MAML adjusts its model parameters using a small number of examples from new, unseen classes and then uses these revised parameters to classify test examples from these classes. We set different inner update learning rates for inner and outer loops in our implementation.As shown in Table~\ref{tab:maml}, the network performs best with an inner learning rate of 0.1. Our model outperforms MAML, achieving a 1.8\% increase in F1 score and a 3.5\% increase in accuracy. 
Although the results obtained with MAML can be improved, MAML's performance is notably hindered by its sensitivity to learning rates. Specifically, higher learning rates can cause inconsistent gradient directions across tasks, while lower rates may lead to underfitting, as MAML struggles to adapt across different task domains\cite{antoniou2018train}. These challenges are worsened by MAML's issues with training instability and inefficient batch normalization, limiting its generalization capabilities. Our method offers a more robust and stable alternative to few-shot learning tasks.



\begin{table}[t]
    \centering
    \caption{Performance evaluation of the MAML algorithm across different shots value.}
    \label{tab:maml}
    \begin{tabular}{|c|c|c|c|c|}
        \hline
        \textbf{Shot} & \textbf{F1-Score} & \textbf{Accuracy}& \textbf{std} & \textbf{CI}\\
        \hline
        1 & 70.3\% & 70.2\% & 0.95 & 0.11 \\
        \hline
        5 & 73.5\% & 73.9\% & 0.21 & 0.07 \\
        \hline
        7 & 74.9\% & 75.4\% & 0.28 & 0.05 \\
         \hline
        10 & 75.0\% & 76.1\% & 0.19 & 0.05 \\
         \hline
    \end{tabular}
    \vspace{-1.5em}
\end{table}

\section{Related Works}
\textbf{Cross-Domain} challenges in the field of medical applications are critical and need to be addressed. This involves leveraging diverse datasets to enhance model robustness and applicability in cross-domain scenarios. Previous studies \cite{zhou2023ensemble,placido2023deep,zhang2021deep} showcase research in this domain, exploring various methods suitable for achieving this goal.
Additionally, recent studies \cite{gadaleta2023prediction,arnaout2021ensemble,wang2021inter,chen2020unsupervised}, further exemplify the methodologies discussed in the context of arrhythmia classification, investigating specific challenges and solutions in this area.
 However, a common limitation is the limited availability of training samples.
This scarcity of data necessitates innovative approaches to training models effectively with a limited number of data. 

\textbf{Few-Shot Learning} has seen recent advancements that present a promising solution to the issue of limited data availability in medical applications.
As demonstrated in \cite{zhu2022enhancing,rahmani2023meta,wang2022high,cui2020unified,feng2021interactive,liu2021metaphys}, research in this area has further explored its versatility and effectiveness across various medical applications.
In the specific field of ECG classification, Siamese networks have been effectively applied to tackle data scarcity. The study in \cite{gupta2021similarity} utilized Siamese Convolutional Neural Networks, outperforming traditional methods like Dynamic Time Warping and LSTM-FCN. Another study, \cite{li2021one}, employed one-dimensional Siamese networks for ECG diagnosis, focusing on feature extraction via shared-weight CNNs and K-nearest similarity judgment for model testing. Additionally, \cite{chen2022meta} combined meta-learning and transfer learning to expedite learning in ECG arrhythmia detection with limited data availability.

\textbf{Transformer} have significantly impacted the medical field,  enhancing the predictive power and interpretability of deep learning algorithms across various tasks, including medical image classification \cite{ai2020correlation,lu2021smile,shome2021covid} and segmentation\cite{chen2021transunet,hatamizadeh2022unetr,karimi2021convolution}. Their capacity to recognize complex patterns in time series data is particularly important in the healthcare industry; this is demonstrated by their application in the analysis of ECG signals to identify and classify arrhythmias. To identify arrhythmias from ECG signals effectively, authors in~\cite{hu2022transformer} develop a vision transformer model named ECG DETR  with a dilated residual network. The DiResViT model is introduced by \cite{pratiher2022dilated}, which combines dilated convolutions with vision transformers to detect atrial fibrillation using time-frequency representations of ECG signals. For enhanced ECG signal analysis and arrhythmia classification, authors in~\cite{che2021constrained} combine CNNs and a transformer model with a unique link constraint.

Building on these works, our work employs TTA in combination with transformers and convolutional networks to address cross-domain challenges in ECG classification. By leveraging deep learning features without manual feature extraction, we aim to improve model robustness and performance in detecting AF, showing the effectiveness of our approach in diverse and challenging real-world scenarios.

\section{Conclusion}
In this study, we employed TTA to distinguish ECG signals of normal individuals from patients with atrial fibrillation, utilizing a cross-domain scenario where the training data from a public dataset and the test data collected independently exhibit significant differences. By leveraging a previously published neural network that integrates transformers for ECG signal encoding and convolutional layers for spectrogram feature extraction, we effectively combined the outputs of these encoders to classify the input signals. Our approach achieved a substantial improvement, reaching an F1 score of 76.8\% and accuracy of 79.6\%.

Additionally, we compared TTA with meta-learning methods, DL and  MAML, using the same training and test datasets. Our findings indicate that TTA can achieve results comparable to or better than those obtained with MAML. Furthermore, TTA enhanced the model's robustness against perturbations in the input signal, demonstrating its efficacy in addressing the cross-domain problem. These results underscore the potential of TTA as a solution for improving model performance and robustness in diverse and challenging real-world scenarios.





\printbibliography


\end{document}